\documentclass[
aps,superscriptaddress,
twocolumn,prl,
showpacs,preprintnumbers,nofootinbib,
amsmath,amssymb,
aps,
superscriptaddress]{revtex4-1}
\usepackage{Preamble}
\makeatletter
\AtBeginDocument{\let\includegraphics\saved@includegraphics}

\newcommand{\bea}{\begin{eqnarray}}
\newcommand{\eea}{\end{eqnarray}}

\begin{document}

\title{Testing mirror symmetry in the Universe with LIGO-Virgo black-hole mergers}

\author{Juan Calder\'{o}n~Bustillo}
	\affiliation{Instituto Galego de F\'{i}sica de Altas Enerx\'{i}as, Universidade de
Santiago de Compostela, 15782 Santiago de Compostela, Galicia, Spain}
	\affiliation{Department of Physics, The Chinese University of Hong Kong, Shatin, N.T., Hong Kong}
\author{Adrian del Rio}
\affiliation{Departamento de F\'isica Te\'{o}rica and IFIC, Universitat de Valencia-CSIC. Dr. Moliner 50, 46100, Burjassot (Valencia), Spain.}
\affiliation{Departamento de Matem\'{a}ticas, Universidad Carlos III de Madrid. Avda. de la Universidad 30, 28911 Legan'{e}s, Spain.}
\author{Nicolas Sanchis-Gual}
\affiliation{Departamento de Astronom\'{i}a y Astrof\'{i}sica, Universitat de Val\`{e}ncia.
Dr. Moliner 50, 46100, Burjassot (Val\'{e}ncia), Spain}

\author{Koustav Chandra}
\affiliation{Institute for Gravitation and the Cosmos, Department of Physics, Pennsylvania State University, University Park, PA 16802, USA}
\affiliation{Department of Physics, Indian Institute of Technology Bombay, Powai, Mumbai, Maharashtra 400076, India}
  
\author{Samson H. W. Leong}
	\affiliation{Department of Physics, The Chinese University of Hong Kong, Shatin, N.T., Hong Kong}

\begin{abstract}

Certain precessing black-hole mergers produce gravitational waves with net circular polarization, understood as an imbalance between right- and left-handed amplitudes. According to the Cosmological Principle, such emission must average to zero across all binary mergers in our Universe to preserve mirror-reflection symmetry at very large scales. We present a new independent gravitational-wave test of this hypothesis. Using a novel observable based on the Chern-Pontryagin pseudo-scalar, we measure the emission of net circular polarization across 47 black-hole mergers recently analyzed by~\cite{NRSurCatalog} with a state-of-the art model for precessing black-hole mergers in General Relativity. The average value obtained is consistent with zero. Remarkably, however, we find that at least $82\%$ of the analysed sources must have produced net circular polarization. Of these, GW200129 shows strong evidence for mirror asymmetry, with a Bayes Factor of 12.6 or, equivalently, $93.1\%$ probability.  We obtain consistent (although stronger) results of $97.5\%$ and $94.3\%$ respectively using public results on this event from~\cite{Hannam_nature_precession} and performing our own parameter inference. This finding further implies evidence of  astrophysical sources that can spontaneously emit circularly polarized photons by quantum effects. Forthcoming black-hole merger detections will enable stronger constraints on large-scale mirror asymmetry and the Cosmological Principle.

\end{abstract}

\maketitle


\noindent\textit{\textbf{Introduction}.} The gravitational-wave (GW) interferometers, Advanced LIGO~\citep{LIGOScientific:2014pky} and Advanced Virgo~\citep{VIRGO:2014yos}, now joined by KAGRA~\citep{KAGRA:2018plz}, have made the detection of compact binary mergers almost routine. In less than a decade, these detectors have reported $\sim \mathcal{O}(90)$  observations~\cite{LIGOScientific:2018mvr,abbott2021gwtc2,abbott2021gwtc3}, providing unprecedented knowledge on  how black holes (BHs) and neutron stars merge in astrophysics, their populations~\cite{Populations_GWTC3}, novel observations of gravitational phenomena~\cite{Vijay_GWKick,Hannam_nature_precession,Kick_GW190412} and  first tests of General Relativity (GR) in the strong-field regime~\cite{Abbott:2016blz, LIGOScientific:2019fpa,GWTC3-TGR,Isi2019_nohair,Bustillo2021,Siegel2023}.

GW observations have also enabled qualitatively new studies of the Universe at large scales by offering a novel, model-independent measurement of the Hubble constant~\cite{H0_nature_lvk, Abbott_2021}. This parameter quantifies the expansion rate of our Universe, and is one of the key observations supporting the standard model of Big Bang cosmology. 
In recent years, however, a serious tension has arisen between different independent measurements of this parameter~\cite{Abdalla:2022yfr,Perivolaropoulos:2021jda}. While systematic errors are still not entirely discarded, the persistence of this and other issues after several years of improved accuracy is starting to put into question the standard cosmological model.
A statistical analysis involving many GW events across the Universe could shed light on some of its underlying global assumptions, such as the Cosmological Principle. 
This is the assumption that, when viewed at sufficiently large scales, our Universe should look homogeneous and isotropic. While observations of the large scale structure~\cite{eBOSS:2020yzd} and anisotropies of the Cosmic Microwave Background~\cite{Planck:2018nkj, Planck2018vyg} support this assumption for $\gtrsim100$ Mpc with great accuracy, various  findings  have recently questioned the validity of this hypothesis~\cite{Abdalla:2022yfr, Perivolaropoulos:2021jda, Aluri:2022hzs}, and have therefore revived interest in this topic. 
Despite the limited number of events, GW catalogs have already made possible some preliminary tests of homogeneity and isotropy in our Universe finding no evidence for violations of these. Such studies target the \textit{extrinsic source parameters}, namely the sky-location ~\cite{Essick2023_distributions} and orientation of the source with respect to the observer \cite{Max_directions,Salvo_orientations}.
In sharp contrast, in this article we describe and demonstrate an independent probe of the Cosmological Principle by testing instead the handedness of the astrophysical BH mergers detected by LIGO-Virgo, which is an \textit{intrinsic} property solely determined by the masses and spins \footnote{We restrict here to quasi-circular BBH mergers, ignoring orbital eccentricity.}.

If our Universe is homogeneous and isotropic at very large scales, then, in particular, it should be statistically invariant under any global mirror transformation. Any evidence of the contrary in large-scale structure observations would violate the Cosmological Principle. 
For isolated astrophysical sources in GR, mirror asymmetry can be quantified through the so-called Chern-Pontryagin scalar. In a spacetime $(M,g_{ab})$, with Riemann curvature tensor $R_{abc}^{\quad\, d}$, the Chern-Pontryagin is the integral $\int_M \dd^4x \sqrt{-g} R^{abcd}{^*R}_{abcd}$, where $*$ represents the Hodge dual, ${^*R}_{abcd}=\frac{1}{2}\epsilon_{abef} R^{ef}_{\quad cd}$. This curvature integral is a (dimensionless) pseudo-scalar and, consequently, flips sign under a mirror transformation. Therefore, it necessarily vanishes if the spacetime $(M,g_{ab})$ has mirror symmetry. A non-zero value can be used to measure mirror asymmetry in astrophysical sources. The higher the Chern-Pontryagin is, the higher the mirror asymmetry of the spacetime will be. \textit{As a coordinate invariant, it is furthermore an observer-independent measure}.

It can be shown that the Chern-Pontryagin is different from zero {if and only if} the (asymptotically flat) spacetime $(M,g_{ab})$ admits a net flux of circularly polarized gravitational radiation, $V_{{\rm GW}}\neq 0$, where~\cite{dRetal20, dR21}:

\bea
V_{{\rm GW}}&=& \int_{0}^{\infty} d \omega \,\omega^{3} \sum_{\ell=2}^{\infty}\sum_{m=-\ell}^{+\ell} \label{VGW}\\ 
&&\quad\left[\bigl|\tilde h^{+}_{\ell m}(\omega)-i \tilde h^{\times}_{\ell m}(\omega)\right|^{2}
-\left|\tilde h^{+}_{\ell m}(\omega)+i \tilde h^{\times}_{\ell m}(\omega)\right|^{2}\bigl] \, , \nonumber 
\eea
can be understood as the gravitational analogue of the electromagnetic Stokes $V$ parameter.  Above, $\tilde h^{+,\times}_{\ell m}(\omega)$ denote the  Fourier transforms of the real and imaginary parts of the  GW strain multipoles, $h_{lm}(t) = h^+_{lm}(t) -i h^{\times}_{lm}(t)$ in a generic  frame, expressed in a spin -2 spherical harmonic basis.

The combinations $\tilde h^{+}_{\ell m}-i \tilde h^{\times}_{\ell m}$ and $\tilde h^{+}_{\ell m}+i \tilde h^{\times}_{\ell m}$ denote a left-handed  and right-handed  circularly polarized wave mode, respectively\footnote{Alternatively, $\tilde h^{+}_{\ell m}-i \tilde h^{\times}_{\ell m}$, $\tilde h^{+}_{\ell m}+i \tilde h^{\times}_{\ell m}$ represent the multipoles of fields with  spin-weight $-2$ and $2$, respectively.}. 
It is easy to see from simple considerations that mirror symmetry implies $V_{GW}=0$, i.e. null net circular polarization. An astrophysical system with mirror symmetry, such as the merger of two spinning black holes with parallel spins, always admits a frame where $\tilde h^+_{\ell \, -m}=(-1)^\ell  \tilde h^+_{\ell m}$, $\tilde h^{\times}_{\ell \, -m}=-(-1)^\ell  \tilde h^{\times}_{\ell m}$. Then, although a particular GW mode $(\tilde h^{+}_{\ell m},\tilde h^{\times}_{\ell m})$ may be circularly polarized, in the sense  $\left|\tilde h^+_{\ell m}-i \tilde h^{\times}_{\ell m}\right|^2-\left|\tilde h^{+}_{\ell m}+i \tilde h_{\ell m}^{\times}\right|^2\neq 0$, the  mode $(\tilde h^+_{\ell \, -m},\tilde h^{\times}_{\ell\, -m})$ cancels this contribution out in the sum above. In other words, if this astrophysical source  generates a circularly polarized flux in a given direction of the sky, observers in the opposite direction will receive exactly the same flux but with opposite circular polarization.  Integrated over the full sphere yields null net handedness. Thus, an imbalance in Eq~\eqref{VGW} between right- and left-handed modes can only be  obtained when there is no mirror symmetry.

Individual astrophysical sources can generate a greater abundance of right-handed over left-handed GWs, or viceversa: $V_{\rm GW}\neq 0$. Since $V_{\rm GW}$ flips sign under a mirror transformation (left-handed modes turn into right-handed modes and viceversa), these sources break mirror symmetry. If the Cosmological Principle holds, however, one should recover mirror-reflection symmetry at very large scales when averaging across all GW sources in our Universe, i.e. $\langle V_{\rm GW} \rangle= 0$. Therefore, GW catalogs can be leveraged to constrain mirror asymmetry in our Universe using $V_{{\rm GW}}$.

As a (pseudo)scalar derived from the Chern-Pontryagin, $V_{{\rm GW}}$ remains invariant under coordinate transformations that preserve the frame's handedness. Consequently, $V_{{\rm GW}}$ is a real number intrinsic to the astrophysical source,  independent of the frame. This allows a test of the Cosmological Principle which is independent of the source's orientation relative to the observer, in contrast to \cite{Essick2023_distributions,Max_directions,Salvo_orientations}.

Binary black holes (BBHs) can produce this imbalance in the handedness of GWs in our universe\footnote{Some modified theories of gravity introduce birefringence effects, i.e. differences in the amplitude of the two  circular polarizations during propagation that can be probed both through gravitational-wave \cite{Wang2022, Ng2023, Callister_birref} and electromagnetic methods \cite{Boudet:2022wmb, Boudet:2022nub, Bombacigno:2022naf}. While the observable (\ref{VGW}) can also be computed within those theories, the present work is concerned with standard GR, where mirror asymmetries are sourced by potential imbalances in BBH populations and not by deviations from GR. }..  However, as elaborated in~\cite{dRetal20, sanchis2023precessing}, mirror asymmetry requires the BH spins to be misaligned not only with respect to the orbital angular momentum (thus leading to orbital precession) but also between them. Therefore, BBHs can only emit such a net flux of circularly polarized radiation (Eq~\eqref{VGW}) in certain precessing configurations. 
The detection of precessing systems itself is far from common, with only one event to date being claimed as precessing, namely GW200129~\cite{Hannam_nature_precession}. This is because current matched-filter searches for BBHs still omit orbital precession in their templates~\cite{abbott2021gwtc3, Messick:2016aqy, Usman:2015kfa, Davies2020}, which greatly damages their sensitivity to these systems~\cite{Bustillo:2016gid, Chandra2020_Nuria, Harry:2016ijz}. Moreover, BBH-spin configurations are influenced by astrophysical formation channels~\cite{Mapelli2020_Summary}: while dynamical formation channels favour isotropic spin distributions~\cite{Sigurdsson1993, PortegiesZwart2000, Rodriguez2016}, common evolution scenarios favour aligned spins~\cite{Tutukov1993, Belczynski2016}.
Do mirror-asymmetric BBH exist in our Universe, or are they just unphysical, idealized scenarios?\\

\begin{figure}[t!]
\begin{center}
\includegraphics[width=0.5\textwidth]{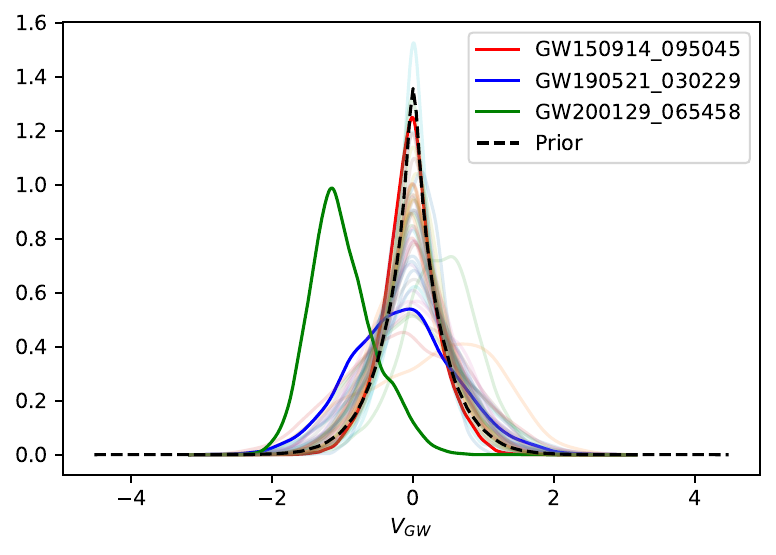}
\caption{\textbf{Posterior probability distribution for the gravitational Stokes parameter (\ref{VGW}) for the 47 events considered in this study}. The black curve denotes the prior distribution. Most of the posteriors are rather non-informative. We highlight the only event displaying strong evidence for non-zero $V_{\rm GW}$, namely GW200129, together with GW190521 and GW150914.}
\label{fig:posterior_deltaQJ_all}
\end{center}
\end{figure}

\begin{figure}[t!]
\begin{center}
\includegraphics[width=0.5\textwidth]{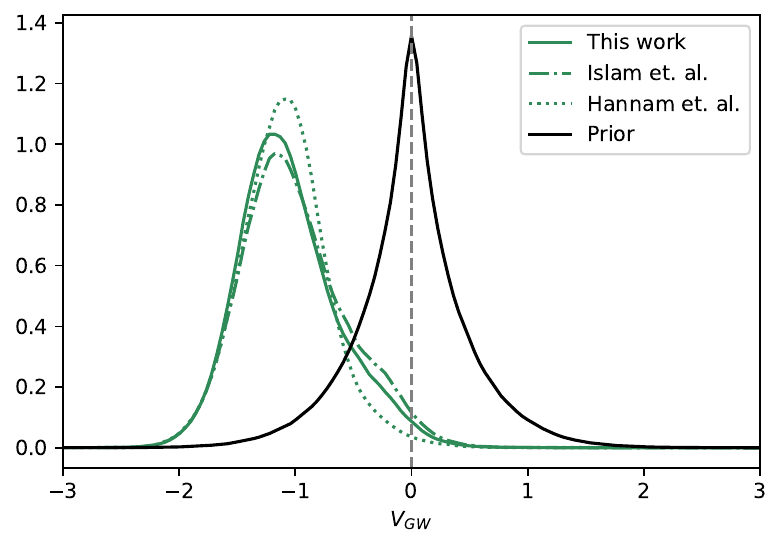}
\caption{\textbf{Posterior probability for the gravitational Stokes parameter $V_{\rm GW}$ for GW200129 according to various analyses}. The solid, dot-dashed and dotted green lines show, respectively, the posterior distributions obtained from our own analysis and those using the samples released by~\citet{NRSurCatalog} and~\citet{Hannam_nature_precession}. We show the prior distribution in black.}
\label{fig:posterior_deltaQJ_200129}
\end{center}
\end{figure}

\begin{figure}[t!]
\begin{center}
\includegraphics[width=0.5\textwidth]{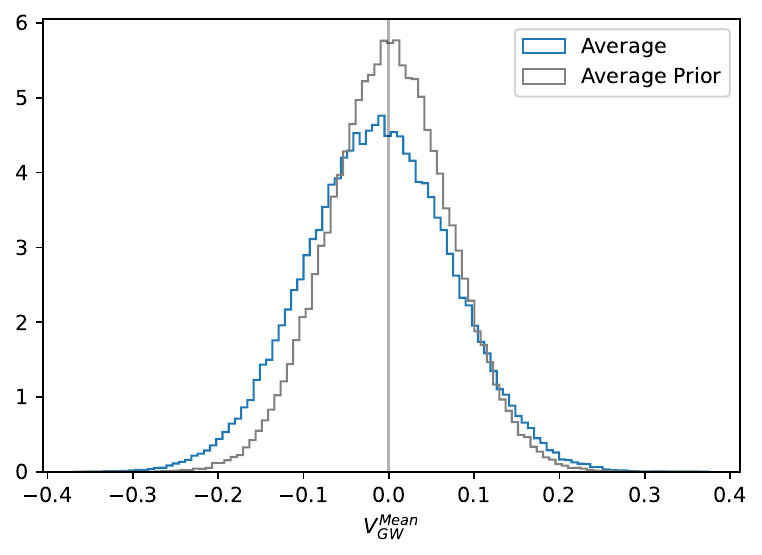}
\caption{\textbf{Posterior probability distribution (in blue) for the average value of $V_{\rm GW}$ across the 47 events considered in this study}. The grey curve denotes the prior distribution. The posterior distribution, while minimally informative, is strongly consistent with zero. Removing GW200129 from the observation set makes the posterior non-informative, following the prior.}
\label{fig:posterior_deltaQJ_mean}
\end{center}
\end{figure}

\noindent\textit{\textbf{Mirror asymmetry in single LIGO-Virgo events}}. We use the posterior distributions for the masses and spins of 47 BBH mergers recently analysed by~\citet{NRSurCatalog}. This analysis uses the state-of-the-art waveform model, \texttt{NRSur7dq4}, which is calibrated directly against numerical relativity simulations obtained by~\citet{SXSCatalog}. \footnote{Due to its direct calibration to numerical relativity, this model naturally includes all the corresponding physics, including the asymmetry between positive and negative $m$  modes characteristic of precessing systems. Other phenomenological models have  been recently proposed in \cite{PhysRevD.109.024061,thompson2023phenomxo4a}. The relevance of  $m$ asymmetries in GW observations has been scrutinized in \cite{kolitsidou2024impact}.  }. The posterior samples are available at~\citet{IslamSamples}. From these, we obtain posterior distributions for $V_{\rm GW}$ through Eq.~\eqref{VGW}, generating the GW modes $h_{\ell m}$ (up to $\ell=4$) using the \texttt{NRSur7dq4} model. 

Figure~\ref{fig:posterior_deltaQJ_all} shows the posterior probability distribution $p(V_{\rm GW})$ for all the 47 events, together with the corresponding prior $\pi(V_{\rm GW})$, shown in black. The latter is induced by the priors on the BBH parameters imposed by~\cite{NRSurCatalog}, described in Appendix II. Among all events, we find that only GW200219 shows a posterior distribution (in green) that clearly deviates from zero, peaking around $V_{\rm GW} = -1$ and with $V_{\rm GW} = 0$ laying at the $98th$ percentile. We also highlight the posterior for the first detection, GW150914~\cite{GW150914_D}, which closely follows the prior, and GW190521, for which~\citet{GW190521D} found mild evidence for precession. 

For each event, we compute the evidence for non-zero $V_{\rm GW}$ through the Savage-Dickey ratio\footnote{See Appendix III.}, given by ${\cal{B}}^{V_{\rm GW}\neq 0}_{V_{\rm GW}=0} = \pi(0)/p(0)$. The corresponding values are reported in Fig.~\ref{fig:Z_ind} (Appendix I) for all events. First, we find that all but one of the events yield ${\cal{B}}^{V_{\rm GW}\neq 0}_{V_{\rm GW}=0} > 1$. Among these GW200219 outstands as the only one yielding strong evidence for $V_{\rm GW} \neq 0$, with ${\cal{B}}^{V_{\rm GW}\neq 0}_{V_{\rm GW}=0} = \pi(0)/p(0) = 12.7$ or, equivalently, $92.7\%$ probability.~\citet{Hannam_nature_precession} also recently analysed this event, finding strong evidence for orbital precession~\cite{Hannam_nature_precession}. Using their posterior samples~\cite{HannamSamples} and the corresponding induced prior, we find  ${\cal{B}}^{V_{\rm GW}\neq 0}_{V_{\rm GW}=0} = 40$ (or $97.5\%$ probability). Finally, performing our own parameter inference (see Appendix II for details), we find an intermediate result of ${\cal{B}}^{V_{\rm GW}\neq 0}_{V_{\rm GW}=0} = 16.7$ (or $94.3$ probability). The corresponding posterior distributions are shown in Fig.~\ref{fig:posterior_deltaQJ_200129}.\\

\noindent{\textit{\textbf{Ensemble properties: average mirror symmetry and relation to orbital precession}}. Figure~\ref{fig:posterior_deltaQJ_mean} shows the posterior distribution for the average of $V_{\rm GW}$ across all 47 events (blue) together with the corresponding prior. The posterior distribution is consistent with zero, mildly deviating from the prior. We obtain $\langle V_{\rm GW} \rangle= -0.013^{+0.142}_{-0.141}$, quoted as the median value, together with the symmetric $90\%$ credible interval. Consistently, using the Savage-Dickey ratio again, we find no evidence that mirror symmetry is violated, obtaining a relative evidence of 1.27:1 between the $\langle V_{\rm GW} \rangle\neq 0$ and the  $\langle V_{\rm GW} \rangle=0$ hypotheses. We have checked that removing GW200129 from the set of events makes the posterior follow the prior. 

While only GW200129 shows strong evidence for non-zero $V_{\rm GW}$, all but one event presents weak positive evidence. This invites the question of what fraction $\zeta_{V_{\rm GW}}$ of the 47 BBH events must have $V_{\rm GW}\neq 0$, even if we cannot identify them individually. We can compute the posterior distribution for $\zeta_{V_{\rm GW}}$ as $p(\zeta_{V_{\rm GW}}) \propto \prod_{i\in[1,47]} \big{[}{\cal{B}}^{V_{\rm GW}\neq 0}_{V_{\rm GW}=0,i} \zeta_{V_{\rm GW}} + (1-\zeta_{V_{\rm GW}})\big{]}$~\cite{Callister2022_zeta,Observations_proca}. We obtain $\zeta_{V_{\rm GW}} = 0.95^{+0.04}_{-0.13}$, indicating that $82\%$ of the events have $V_{\rm GW}\neq0$, which requires orbital precession, at the $95\%$ credible level (see the discussion section for more details).\\

\begin{figure}[t!]
\begin{center}
\includegraphics[width=0.49\textwidth]{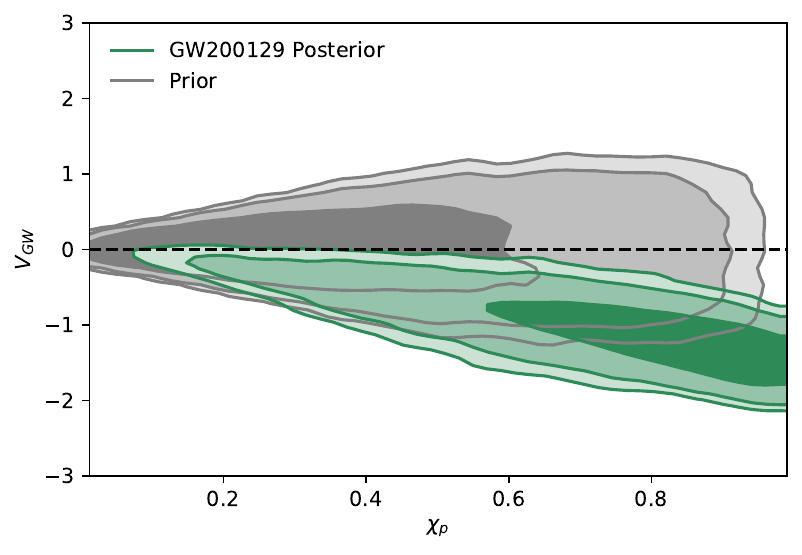}
\caption{\textbf{Gravitational Stokes $V_{\rm GW}$ parameter vs. effective orbital precession parameter $\chi_p$ for GW200129}. We show two-dimensional $69\%,90\%$ and $95\%$ credible contours, together with the corresponding prior.}  
\label{fig:2dplot}
\end{center}
\end{figure}

\begin{figure}[t!]
\begin{center}
\includegraphics[width=0.5\textwidth]{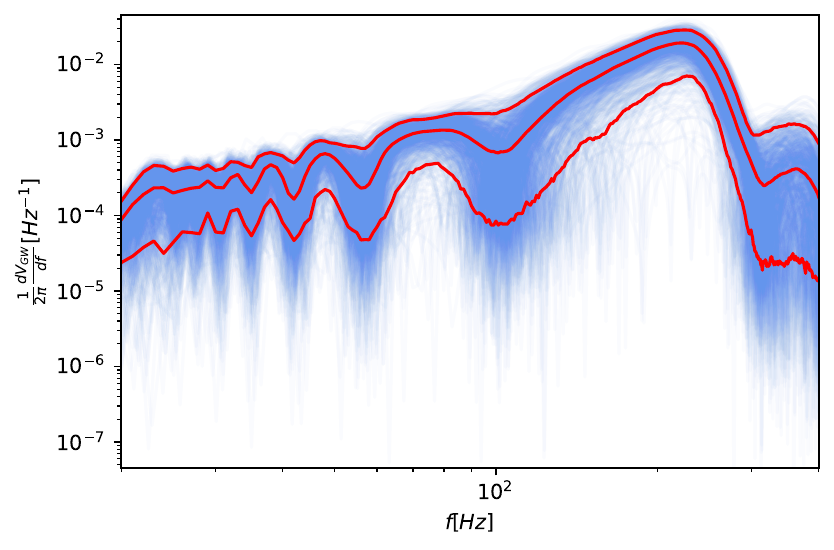}
\caption{\textbf{Spectrum for the circularly polarised emission of GW200129}. Posterior distribution of the absolute value of the integrand of $V_{\rm GW}$ as a function of the frequency. The red lines denote the median and $90\%$ credible intervals. 
}
\label{fig:spectrum}
\end{center}
\end{figure}

\noindent{\textbf{\textit{Beyond orbital precession}}}. 
Given the close relation between orbital precession and $V_{\rm GW}$, it is natural to ask whether a non-zero $V_{\rm GW}$ for GW200129 follows trivially from previous claims of orbital precession in this event.

To answer this question, Fig.~\ref{fig:2dplot} shows the two-dimensional $69\%,90\%$ and $95\%$ credible regions for $V_{\rm GW}$ and the effective-precession spin parameter $\chi_p$~\cite{Schmidt:2014iyl, Hannam:2013oca}, commonly used to parametrise the amount of precession. First, the posterior for GW200129 clearly deviates from the prior. Second, while the prior distribution shows a clear correlation between the value of $\chi_p$ and the maximum allowed value of $V_{\rm GW}$, GW200129 clearly deviates from the prior even when restricting to high values of $\chi_p$, progressively disfavouring $V_{\rm GW}=0$ for increasing $\chi_p$. To quantitatively show this, we compute ${\cal{B}}^{V_{\rm GW}\neq 0}_{V_{\rm GW}=0}$ for restricted intervals of $\chi_p \in [0.7,0.8]$, $\chi_p \in [0.8,0.9]$ and $\chi_p \in [0.9,1]$, respectively yielding 17, 33 and 53. In summary, our finding of mirror asymmetry for GW200129 does not trivially follow from that of orbital precession. Instead, this event belongs to a ``sub-class'' of precessing mergers producing a net emission of circularly polarized GWs.\\

\noindent\textit{\textbf{Spectrum}}. Figure~\ref{fig:spectrum} shows the posterior distribution for the absolute value of the integrand of $V_{\rm GW}$ for GW200129 as a function of the frequency. The red lines denote, for each frequency, median values and $90\%$ credible intervals. We highlight two aspects. First, the highest emission probability occurs at $200$\,Hz,  roughly corresponding to the merger frequency. This is not surprising as it is near a merger when mirror asymmetry becomes stronger. Second, while a lower cutoff of 20\,Hz was used in our calculation of $V_{\rm GW}$,
Fig.~\ref{fig:spectrum} clearly shows that, for such frequencies, the integrand has already decayed nearly two orders of magnitude with respect to the peak value. Therefore, we do not expect the limited length of our waveforms to impact our conclusions.\\

\noindent\textit{\textbf{Discussion}}. Since GR is not a  chiral theory, one may  assume that our Universe should contain the same proportion of ``right-handed''  vs ``left-handed''  BBHs to achieve global mirror symmetry. However, this is not necessarily the case\footnote{An asymmetry can be developed, for instance, from certain initial conditions in the early Universe, or from specific astrophysical formation channels that may favor one particular handedness.}, and this  hypothesis has to be confirmed with observations. This is crucial, as any significant deviation  would lead to a direct violation of the Cosmological Principle.
GW astronomy has now compiled a good sample of $\sim$ 90 events, and is starting to have the potential to test this hypothesis. Currently observed BBHs range in distances up to 10 Gpc, providing valid probes of large-scale phenomena. Using these, we have designed a novel test based on the calculation of the net circular polarization emission from 47 BBHs observed to date,
through the observable $V_{\rm GW}$, derived from the Chern-Pontryagin pseudo-scalar.

We report three main results. First, we find \textit{no evidence that mirror symmetry is broken across the observation set}. Second, we find \textit{strong evidence that GW200129 produced a net emission of circularly polarized GWs} -- with Bayesian evidence that varies between 12.6 and 40, depending on the analysis choices --  with all but one of the remaining events mildly favouring such hypothesis. This result is consistent with the fact that GW200129 has previously been identified as a precessing BBH, which is a necessary condition for a net GW polarization emission~\cite{dRetal20, sanchis2023precessing}. Remarkably, while most posteriors for individual events are rather non-informative and all but one (very mildly) favour the net emission hypothesis, any deviations of the average posterior from the corresponding prior are solely due to GW200129. This situation may change as further GW events, especially strongly precessing events, are detected in current and forthcoming observation runs. Third, we have found that at least $82\%$ of the analysed events must have $V_{\rm GW}\neq 0$, which requires orbital precession. We cannot, however, claim that $82\%$ of the events must be precessing, as none of our competing models is purely restricted to non-precessing systems. Instead, we show that if BBH mergers are split into two physically motivated groups -- one containing only precessing mergers -- most binaries must belong to such group, with the other one containing both aligned-spin and precessing mergers with $V_{\rm GW}=0$.

The existence of mirror asymmetric BBHs like GW200129 is the breeding ground for additional parity-violating processes. As reported in~\cite{dRetal20}, these systems trigger the spontaneous creation of photon pairs with the same helicity by quantum effects~\cite{AdRNS17a,AdRNS17b,AdRNS18a,AdRNS18b}, thus producing an overabundance of photons with one preferred handedness. Integrated across all the events in the Universe, they might impact Cosmological observables, e.g. by introducing a systematic error in measurements of the CMB. 
In this work, a viable astrophysical source for this quantum effect has been identified for the first time, shedding evidence that these sources are viable and potentially frequent in Nature.
Our results can have deep implications in the context of BBH formation channels and their environments. The hints found on the number of precessing events, which can only be confirmed through explicit calculation of Bayesian evidences for such effect}, aligns with existing arguments for dominant dynamical formation channels (although these rather come from claims of orbital eccentricity~\cite{RomeroShaw2022,Gayathri2022_ecc_natastro,RomeroShaw2020_ecc_apjl}); as opposed to isolated evolution, which favours aligned spins typical of isolated BBH formation \cite{Sigurdsson1993,PortegiesZwart2000,Rodriguez2016}. Dynamical formation is characterised by isotropic spin distributions \cite{Tutukov1993,Belczynski2016} and is linked to hierarchical BBH formation in dense environments, commonly invoked to explain the current observations of merging black holes \cite{Carlos_Henry_2404.00720,Mahapatra2024,Barrera_spins,GW190521D,GW190521I} with masses within the Pair-instability supernova gap \cite{Heger:2002by,Woosley2021,Farmer2019_PISN} and the formation of supermassive black holes from stellar-mass ones \cite{Volonteri2010}. While it is known that precession and eccentricity can be mistaken for very large total masses~\cite{Bustillo2021_HeadOn,RomeroShaw2020_ecc_apjl}, most of the studied events have masses low enough that they should be safe from such degeneracy. 
Our study is also limited by the fact that we do not include selection or population effects, which would allow us to conclude not only on the observed set of events but also on the underlying population. In particular, it is known that current searches for BBHs have very limited sensitivity to precessing BBHs as compared to aligned-spin ones, hinting that the fraction of precessing events in the population may be much larger than that in the observation set. Estimating such sensitivities would, however, require computationally intensive injection campaigns~\cite{o3_imbh, Chandra2020_Nuria}, so we leave this is as a potential future line of research.

Finally, our study omits the handedness of sources beyond BBHs as, e.g., neutron stars, which can emit a comparable GW strain~\cite{GW170817_PRL, BNS_NR_Dietrich2015} and can produce non-zero $V_{\rm GW}$ even in the absence of precession, during their post-merger stages. Such signals should be detectable by future detectors like Cosmic Explorer~\cite{CE, CE2}, Einstein Telescope~\cite{ET1, ET2} or NEMO~\cite{NEMO_Ackley2020}. The GW background recently detected by NANOGrav~\cite{NANOGrav:2023gor}, which remains out of the scope of this paper for obvious reasons, may also display some degree of mirror asymmetry and contribute to another frequency band.\\

\noindent\textit{\textbf{Acknowledgements}}. We thank Angela Borchers for her advice on extracting individual modes from NRSur7dq4. We also thank Thomas Dent, Aditya Vijaykumar, Will Farr, Tjonnie Li and Cecilio Garcia-Quiros for useful discussions and comments on the manuscript. JCB is funded by a fellowship from ``la Caixa'' Foundation (ID100010474) and from the European Union's Horizon2020 research and innovation programme under the
Marie Skodowska-Curie grant agreement No 847648. The fellowship code is LCF/BQ/PI20/11760016. JCB is also supported by the research grant PID2020-118635GB-I00 from the Spain-Ministerio de Ciencia e Innovaci\'{o}n and by its Ram\'on y Cajal program (grant RYC2022-036203-I). 
ADR acknowledges support through (i) {\it Maria Zambrano} grant ZA21-048 with reference UP2021-044 from the Spanish Ministerio de Universidades, funded within the European Union-Next Generation EU, and (ii) {\it Atraccion de Talento Cesar Nombela} grant No 2023-T1/TEC-29023, funded by Comunidad de Madrid (Spain). ADR further acknowledges financial support  via the Spanish Grants PID2020-116567GB-C21, funded by MCIN/AEI/10.13039/501100011033, and PID2023-149560NB-C21, funded by MCIU /AEI/10.13039/501100011033/FEDER, UE.
NSG acknowledges support from the Spanish Ministerio de Universidades, through a Mar\'ia Zambrano grant (ZA21-031)
with reference UP2021-044, funded within the European Union-Next Generation EU, and from the Spanish Ministry of Science and Innovation via the Ram\'on y Cajal programme (grant RYC2022-037424-I), funded by MCIN/AEI/ 10.13039/501100011033 and by ``ESF Investing in your future”. NSG is further supported by the Spanish Agencia Estatal de Investigaci\'on (Grant PID2021-125485NB-C21) funded by
MCIN/AEI/10.13039/501100011033 and ERDF A way
of making Europe, 
and by the European Union's Horizon 2020 research and innovation
(RISE) programme H2020-MSCA-RISE-2017 Grant No.
FunFiCO-777740 and by the European Horizon Europe staff exchange (SE) programme HORIZON-MSCA2021-SE-01 Grant No. NewFunFiCO-101086251. KC acknowledges the support through NSF grant numbers PHY-2207638, AST-2307147, PHY-2308886, and PHY-2309064.
We acknowledge the use of IUCAA LDG cluster Sarathi for the computational/numerical work. The authors acknowledge computational resources provided by the CIT cluster of the LIGO Laboratory and supported by National Science Foundation Grants PHY-0757058 and PHY0823459; and the support of the NSF CIT cluster for the provision of computational resources for our parameter inference runs. This material is based upon work supported by NSF's LIGO Laboratory which is a major facility fully funded by the National Science Foundation. The analysed LIGO-Virgo data and the corresponding power spectral densities, in their strain versions, are publicly available at the online Gravitational-Wave Open Science Center~\cite{SoftwareX,OpenDataArxiv}. This research has made use of data or software obtained from the Gravitational Wave Open Science Center (gwosc.org), a service of LIGO Laboratory, the LIGO Scientific Collaboration, the Virgo Collaboration, and KAGRA. LIGO Laboratory and Advanced LIGO are funded by the United States National Science Foundation (NSF) as well as the Science and Technology Facilities Council (STFC) of the United Kingdom, the Max-Planck-Society (MPS), and the State of Niedersachsen/Germany for support of the construction of Advanced LIGO and construction and operation of the GEO600 detector. Additional support for Advanced LIGO was provided by the Australian Research Council. Virgo is funded, through the European Gravitational Observatory (EGO), by the French Centre National de Recherche Scientifique (CNRS), the Italian Istituto Nazionale di Fisica Nucleare (INFN) and the Dutch Nikhef, with contributions by institutions from Belgium, Germany, Greece, Hungary, Ireland, Japan, Monaco, Poland, Portugal, Spain. KAGRA is supported by Ministry of Education, Culture, Sports, Science and Technology (MEXT), Japan Society for the Promotion of Science (JSPS) in Japan; National Research Foundation (NRF) and Ministry of Science and ICT (MSIT) in Korea; Academia Sinica (AS) and National Science and Technology Council (NSTC) in Taiwan. This manuscript has LIGO DCC number P2400018. 

\appendix
\section*{Appendix I: Individual evidences for non-zero $V_{\rm GM}$}

\begin{figure*}[t!]
\begin{center}
\includegraphics[width=1\textwidth]{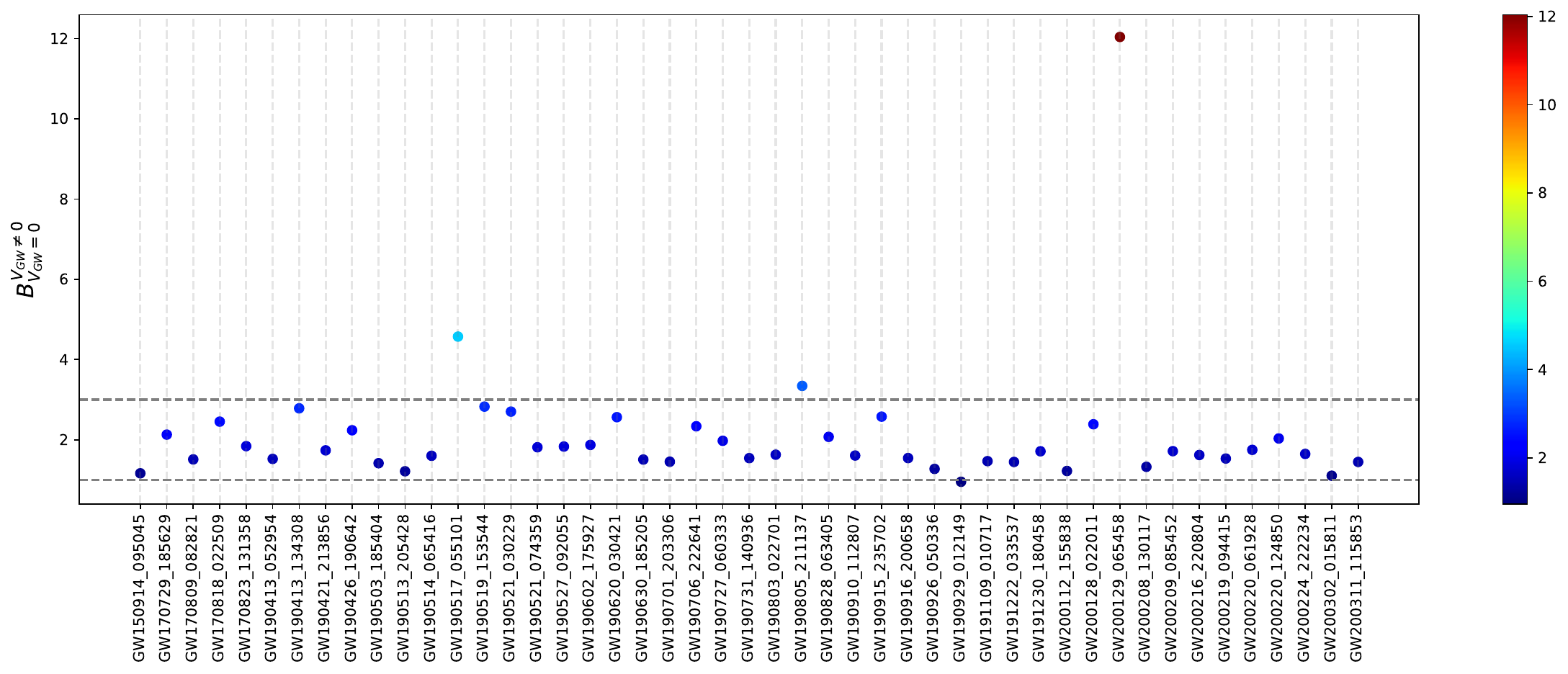}
\caption{\textbf{Evidence for non-zero $V_{\rm GW}$ for each of the analysed events}. For each event, we show the relative Bayes factor for the $V_{\rm GW}\neq 0$ scenario versus the $V_{\rm GW} = 0$ one, according to the posterior parameter samples released by~\cite{NRSurCatalog}.}
\label{fig:Z_ind}
\end{center}
\end{figure*}

Figure~\ref{fig:Z_ind} shows the value of  ${\cal{B}}^{V_{\rm GW}\neq 0}_{V_{\rm GW}=0}$ for each of the 47 BBHs considered in this study, obtained from the posterior samples publicly released by~\cite{NRSurCatalog}. As described in the text, all values are above 1 expect for that for GW190929\_012149. The event with the second highest value, GW190527\_055101, does barely exceed a value of 4, not individually indicative of any strong evidence for the effect we discuss.

\section*{Appendix II: Comparison of parameter Inference Settings}\label{sec:appendix-2}

We conduct Bayesian parameter inference on 8-second data from the pair of Advanced LIGO and the Advanced Virgo detectors around the time of GW200129 and other selected events. Our analysis uses the publicly available software \texttt{Bilby}~\cite{Ashton:2018jfp, Isobel-catalogue} in its parallelizable version \texttt{Parallel Bilby}~\cite{pbilby}. For the GW200129 study, we use the de-glitched data frame for LIGO Livingston as the event coincided with an excess noise caused by an electro-optic modular system~\citep{abbott2021gwtc3} similar to~\citet{Hannam_nature_precession} and~\citet{Islam:2023zzj}. 

As a signal-template model, we use the state-of-the-art model \texttt{NRSur7dq4}~\cite{NRSur7dq4}, which is directly calibrated to numerical-relativity simulations of precessing binary black-hole mergers and includes higher emission modes up to $\ell=4$. We use the standard likelihood function for gravitational-wave transients~\cite{Finn1992, Romano2017} and evaluate it from a fixed minimum frequency of $f_\mathrm{min}=20\,$Hz identical to the reference frequency choice for our analysis. Our priors align with those adopted by the LIGO-Virgo-KAGRA collaboration, except for the bounds in the mass parameter space, which we adapt to the domain of validity of the NRSur7dq4 model. Specifically, we restrict the mass ratio to values greater than 1/6 and constrain the chirp mass from $14\,M_\odot$ to $100\,M_\odot$. We sample the likelihood function across the parameter space using the nested-sampling algorithm \texttt{dynesty}~\cite{Dynesty} and use Eq.~\eqref{VGW} to obtain the gravitational Stokes parameter $V_{\rm GW}$ from the obtained posterior samples for the source.

For comparison, the analysis by~\citet{Hannam_nature_precession} limits the mass ratio between 1/4 and 1, the chirp mass between $14.5\,M_\odot$ and $49\,M_\odot$ and the detector frame total mass to be above $68\,M_\odot$. Also, they used a reference frequency of 50\,Hz and a power law luminosity distance $D_L^2$ prior default in LALInference~\citep{lalinference_o2}. The latter is in contrast with our luminosity-distance prior, which corresponds to a uniform merger rate in the source’s co-moving frame for a cold dark matter cosmology with $H_0 = 67.9\ \textrm{km/s/Mpc}$ and $\Omega_m = 0.3065$, similar to that used for LVK's and~\citet{Islam:2023zzj} analysis. We believe this is the main reason behind the slightly different results shown in Fig.~\ref{fig:posterior_deltaQJ_200129}, as the little ``bump'' of the posterior distribution for $V_{\rm GW}$ around zero in ours and Islam's analyses correlates with a similar feature in the posterior distribution for the luminosity distance, located at low distance values.

The~\citet{Islam:2023zzj} analysis also employs marginally distinct mass prior bounds. They restrict the mass ratio within the range of 1/6 to 1, constrain the chirp mass to fall between $12\,M_\odot$ and $400\,M_\odot$, and set the detector frame total mass in the interval $60\,M_\odot$ to $400\,M_\odot$, while maintaining consistency with other settings employed in our analysis.

\section*{APPENDIX III: Savage-Dickey density ratio for generic sharp nested hypotheses}\label{sec:appendix-3}

Consider two models $H_1$ and $H_0$ described by a set of parameters collectively denoted as $\vec\theta$ and an extra parameter $\varphi$. Next, consider that the $H_1$ model allows all parameters to vary freely over a given range, with prior probability $\pi_1(\vec\theta,\varphi)$ while the model $H_0$ imposes the restriction $\varphi=\varphi_0$. Further, let us assume that the prior probability $\pi_0(\vec\theta)$ for the parameters $\vec\theta$ in the model $H_0$ is equal to the prior probability for $\vec\theta$ in $H_1$, conditional to $\varphi=\varphi_0$. This is, $\pi_0(\vec\theta)=\pi_1(\vec\theta | \varphi=\varphi_0)$. Under these conditions, the Bayes' Factor for the model $H_0$ over $H_1$ is given by the ratio of the marginal posterior and prior distributions for $\varphi=\varphi_0$ in $H_1$, known as the Savage-Dickey ratio. This is:
\begin{equation}
   {\cal{B}}^{0}_{1} = \frac{p_1(\varphi=\varphi_0 | d)}{\pi_1(\varphi=\varphi_0)}, 
\end{equation}
where
\begin{equation}
\begin{aligned}
 & p_1(\varphi = \varphi_0|d) = \int  \delta(\varphi-\varphi_0)p_1(\vec{\theta},\varphi|d) \,\dd\vec{\theta} \,\dd \varphi\, , \\
 & \pi_1(\varphi = \varphi_0) = \int  \delta(\varphi-\varphi_0)\pi_1(\vec{\theta},\varphi) \,\dd \vec{\theta} \,\dd \varphi.
\end{aligned}
\end{equation}
Above, $p_1(\vec\theta|d)$ denotes the posterior probability for the parameters $\vec\theta$, given the data $d$, under the model $H_1$. In general, the posterior probability distribution for a set of parameters $\vec\theta$ is given by 
\begin{equation}
   p(\vec{\theta}|d) = \frac{p (\vec\theta)p(d|\vec{\theta})}{p_1(d)} \equiv \frac{\pi (\vec\theta){\cal{L}}(\vec{\theta})}{{\cal{Z}}}.
\end{equation}
The term ${\cal{L}}(\vec\theta) \equiv p(d|\vec\theta)$ denotes the probability for the data $d$ given the parameters $\vec\theta$ (also simply known as the ``likelihood'' for $\vec\theta$) and the term ${\cal{Z}}$ denotes the evidence for the model $H$, given by
\begin{equation}
 {\cal{Z}} = \int \pi (\vec\theta){\cal{L}}(\vec{\theta}) \,{\rm d}\vec{\theta}.
\end{equation}

In the main text, we have applied the Savage-Dickey ratio to a nested model $H_0$ defined as the restriction of a larger model $H_1$, with parameters $\vec{\theta}=\{\theta_1,...,\theta_n\}$, through a condition imposed on a function $f(\vec\theta)$ of the parameters $\vec\theta$. In contrast, in all of the cases we have seen in the literature, the Savage-Dickey ratio is applied to models $H_0$ defined through the restriction of one of the parameters $\theta_i$ to a fixed value $\theta_{i,0}$. While it seems natural that the Savage-Dickey ratio should be applicable the situation we consider, we have not been able to find an explicit demonstration in the literature. For this reason, we provide a derivation of it in this Appendix.\\

The relative Bayes' Factor for the model $H_0$ over the model $H_1$ is given by the ratio of the respective Bayesian evidences. This is: 
\begin{equation}
     {\cal{B}}^{0}_{1}  = \frac{{\cal{Z}}_0}{{\cal{Z}}_1}.
\end{equation}
By definition, the evidence for the model $H_0$ is given by:
\begin{equation}
    {\cal{Z}}_0 = \int \pi_0(\vec\theta){\cal{L}}(\vec\theta)\,\dd \vec{\theta}.
\end{equation}
Also by definition, $H_0$ is given by the restriction of $H_1$ to the subset of parameters satisfying $f(\vec{\theta})=0$. Therefore, the prior probability for the parameters $\vec{\theta}$ in $H_0$ is given by $\pi_0(\theta) = \pi_1(\theta | f(\theta)=0)$. 
With this, we get 
\begin{equation}
    {\cal{Z}}_0 = \int \pi_1(\vec{\theta} | f(\vec{\theta})=0){\cal{L}}(\vec\theta)\,\dd \vec{\theta}.
\end{equation}
Applying the Bayes' rule to the first term in the integrand yields
\begin{equation}
    {\cal{Z}}_0 = \int \frac{\pi_1(f(\vec{\theta})=0 | \vec{\theta})\pi_1(\vec{\theta})}{\pi_1(f(\vec{\theta})=0)}{\cal{L}}(\vec\theta) \,\dd\vec{\theta}.
\end{equation}
Next, using the definition for the evidence for $H_1$, we obtain
\begin{equation}
    {\cal{Z}}_0 = {\cal{Z}}_1 \int \frac{\pi_1(f(\vec{\theta})=0 | \vec{\theta})}{\pi_1(f(\vec{\theta})=0)} p_1(\vec{\theta}|d) \,\dd\vec{\theta}.
\end{equation}
Finally, noticing that $\pi_1(f(\vec{\theta})=0 | \vec{\theta}) = \delta(f(\vec\theta)-0)$, we arrive to the expression
\begin{equation}
    {\cal{Z}}_0 = {\cal{Z}}_1 \frac{p_1(f(\vec{\theta})=0|d)}{\pi_1(f(\vec{\theta})=0)},
\end{equation}
where
\begin{equation}
\begin{aligned}
 & p_1(f(\vec{\theta})=0|d) = \int  \delta(f(\vec\theta)-0)p_1(\vec{\theta}|d) \,\dd\vec{\theta}\, 
\end{aligned}
\end{equation}
denotes the marginal posterior distribution for $f(\theta)=0$ within the model $H_1$.

\bibliography{boson_pop.bib,psi4_observation}

\end{document}